\documentclass[11pt]{article}
\usepackage{amssymb}
\usepackage{amsmath}
\usepackage{amscd}
\usepackage{latexsym}

\catcode`@=11
\renewcommand{\section}{\@startsection{section}{1}{0pt}{\medskipamount}
{\medskipamount}{\large\bf}}
\numberwithin{equation}{section}
\newcounter{saveeqn}
\newcommand{\alpheqn}{\setcounter{saveeqn}{\value{equation}}%
\stepcounter{saveeqn}%
\setcounter{equation}{0}%
\renewcommand{\theequation}{%
              \mbox{\arabic{section}.\arabic{saveeqn}\alph{equation}}}}%
\newcommand{\reseteqn}{\setcounter{equation}{\value{saveeqn}}%
\renewcommand{\theequation}{\arabic{section}.\arabic{equation}}}
\catcode`@=12
\def\a{\alpha}
\def\b{\beta}
\def\g{\gamma}

\def\de{\delta}
\def\eps{\epsilon}

\def\vk{\varkappa}

\def\la{\lambda}

\def\p{\phi}
\def\vp{\varphi}

\def\om{\omega}

\def\sn{\mbox{sn}}

\newcommand{\R}{\mathbb R}
\newcommand{\Z}{\mathbb Z}
\newcommand{\NN}{\mathbb N}

\newcommand{\Acal}{{\cal A}}

\newcommand{\Ecal}{{\cal E}}
\newcommand{\gfrak}{{\mathfrak g}}

\newcommand{\Fcal}{{\cal F}}
\newcommand{\mfrak}{{\mathfrak m}}
\newcommand{\hfrak}{{\mathfrak h}}

\def\im{\textrm{i}}

\def\N2{$N{=}2$}

\def\diff{\textrm{d}}
\def\tr{\textrm{tr}}
\def\sfrac#1#2{{\textstyle\frac{#1}{#2}}}
\def\>{\rangle}
\def\<{\langle}
\def\+{\dagger}
\def\={\ =\ }
\def\und{\qquad\textrm{and}\qquad}

\def\ph{\hat{\phi}}

\begin{document}

\begin{titlepage}
\setcounter{page}{0}
\begin{flushright}
ITP--UH--05/09\\
\end{flushright}

\vskip 2.0cm

\begin{center}

{\Large\bf 
Instantons and Yang-Mills Flows on Coset Spaces 
}

\vspace{12mm}

{\large Tatiana~A.~Ivanova${}^*$,\ Olaf~Lechtenfeld${}^\+$,\ Alexander~D.~Popov${}^*$ \ and\ Thorsten~Rahn${}^\+$
}
\\[8mm]
\noindent ${}^*${\em Bogoliubov Laboratory of Theoretical Physics, JINR\\
141980 Dubna, Moscow Region, Russia}\\
{Email: ita, popov@theor.jinr.ru}
\\[8mm]
\noindent ${}^\+${\em Institut f\"ur Theoretische Physik,
Leibniz Universit\"at Hannover \\
Appelstra\ss{}e 2, 30167 Hannover, Germany }\\
{Email: lechtenf, rahn@itp.uni-hannover.de}

\vspace{12mm}

\begin{abstract}
\noindent
We consider the Yang-Mills flow equations on a reductive coset space $G/H$ and 
the Yang-Mills equations on the manifold $\R\times G/H$. On nonsymmetric coset
spaces $G/H$ one can introduce geometric fluxes identified with the torsion 
of the spin connection. The condition of $G$-equivariance imposed on the gauge 
fields reduces the Yang-Mills equations to $\p^4$-kink equations on~$\R$. 
Depending on the boundary conditions and torsion, we obtain solutions to the 
Yang-Mills equations describing instantons, chains of instanton-anti-instanton 
pairs or modifications of gauge bundles. For Lorentzian signature on 
$\R\times G/H$, dyon-type configurations are constructed as well. We also 
present explicit solutions to the Yang-Mills flow equations and compare them 
with the Yang-Mills solutions on~$\R\times G/H$.

\end{abstract}

\end{center}
\end{titlepage}

\section{Introduction and summary}

\noindent
The Yang-Mills equations in more than four dimensions naturally appear
in the low-energy limit of superstring theory. Furthermore, natural BPS-type
equations for gauge fields in dimensions $d>4$, introduced in~\cite{CDFN, W1},
also appear in superstring compactifications as the conditions for the
survival of at least one supersymmetry in low-energy effective field
theory in four dimensions~\cite{GSW}. For non-Abelian gauge theory on a
K\"ahler manifold the most natural BPS condition lies in the
Donaldson-Uhlenbeck-Yau equations~\cite{Don, UY}. While the criteria for the
existence of solutions to all these equations are by now well-understood
~\cite{Don, UY, Car}, in practice it is usually quite difficult to find their
explicit form. Some solutions in $\R^d$ were found e.g.~in~\cite{group1,group2}
but have infinite action for $d>4$. One possibility for obtaining finite-action
solutions to the Yang-Mills equations in higher dimensions is to introduce a 
noncommutative deformation of this field theory~\cite{group3}. With the help 
of a Moyal deformation some finite BPS and non-BPS solutions have been produced
e.g.~in~\cite{group4}-\cite{group6}, alongside with their brane interpretation.

Let us describe another way to generate finite-action solution, by removing
the origin from~$\R^d$. Then, the two metrics
\begin{equation}\label{1.1}
\diff s_1^2\=\diff r^2 + r^2\diff\Omega^2_{d-1}
\end{equation}
and
\begin{equation}\label{1.2}
\diff s_2^2\=r^{-2}\left (\diff r^2 + r^2\diff\Omega^2_{d-1}\right )
\ =:\ \diff\tau^2 + \diff\Omega^2_{d-1}
\end{equation}
on $\R^d-\{0\}\cong\R\times S^{d-1}$ are conformally equivalent. 
For $d=4$, the condition of SO(4)-invariance reduces the Yang-Mills equations 
on $\R\times S^3$ with the metric (\ref{1.2}) to the $\p^4$-kink equation for
a scalar field $\p$ depending on $\tau\in\R$ (see e.g.~\cite{popov1}). 
Inserting the kink solution into the gauge-field ansatz yields precisely the
BPST one-instanton configuration~\cite{BPST}, which can be extended from 
$\R\times S^3$ to $\R^4$ or $S^4$ due to the conformal invariance of Yang-Mills
theory in four dimensions. In~\cite{IvLe} this setup was carried to spaces
$\R\times G$, with $G$ being a compact semisimple Lie group and 
$\diff\Omega^2_{d-1}$ in~(\ref{1.2}) denoting the standard left-invariant 
metric on~$G$. It was shown that the Yang-Mills equations on $\R\times G$ 
for an $Ad_G$-invariant gauge potential also reduce to kink equations, whose 
solutions give instantons on $\R\times G$ with finite action and topological 
charge.

In this paper, we generalize the results of~\cite{popov1, IvLe} on symmetric 
Yang-Mills solutions on $\R\times S^2$, $\R\times S^3$ and $\R\times G$ to any
space of the form $\R\times G/H$, where $G/H$ is a reductive homogeneous space
(coset space). 
First, we construct explicit instanton solutions for the case where $G/H$ is 
compact and symmetric. These configurations describe transitions between two
$G/H$ Yang-Mills vacua (at $\tau=\pm\infty$). Imposing periodicity in $\tau$, 
we also obtain solutions describing chains of instanton-anti-instanton pairs. 
Second, we generate solutions for nonsymmetric coset spaces $G/H$ with 
geometric torsion. These configurations describe modifications of gauge bundles
(cf.~\cite{ KW, LOZ}) over $G/H$, in particular transitions from the vacuum
at $\tau =-\infty$ to a topologically nontrivial Yang-Mills solution on $G/H$ 
at~$\tau =+\infty$. In the special case of $G/H=S^{d-1}$ we arrive at explicit
Yang-Mills solutions on $\R^d-\{0\}$ which, however, cannot be smoothly 
extended to $\R^d$ due to the lack of conformal invariance of the Yang-Mills 
equations in $d>4$ dimensions.

Third, we investigate the Yang-Mills flow equations on coset spaces.
The Yang-Mills flow on a manifold $M$ is a one-parameter family of connections
$\Acal (\tau)$ with $\tau\in\R$, determined by the equation
\begin{equation}\label{1.3}
\frac{\diff}{\diff\tau}\Acal \= -*\nabla (*\Fcal )\ ,
\end{equation}
where $\nabla$ is the covariant derivative, $\Fcal$ is the curvature and
$*$ is the Hodge star operator on $M$~\cite{Don}. At critical points $\tau_0$, 
where $\frac{\diff\Acal}{\diff\tau}|_{\tau_0}=0$, these equations reduce to the
Yang-Mills equations $\nabla(*\Fcal )=0$ on $M$. Mostly studied in the 
literature is the existence, uniqueness and regularity properties of the
Yang-Mills flows in two, three and four dimensions (see e.g.~\cite{Don, HT} 
and references therein). In this paper, we create explicit solutions to the 
Yang-Mills flow equations on reductive coset spaces $G/H$ by using our 
$G$-equivariant ansatz for the gauge potential~$\Acal$. These configurations 
describe modifications of gauge bundles over $G/H$ similar to those obtained 
for the full Yang-Mills solutions on $\R\times G/H$.

\bigskip

\section{From Yang-Mills on $\R\times G/H$ to the kink}

\noindent
{\bf Coset spaces $G/H$.}   
Consider a compact semisimple Lie group $G$ and a closed subgroup $H$ of $G$ 
such that $G/H$ is a reductive homogeneous space (coset space). 
Let $\{I_A\}$ with $A{=}1,\ldots,\,$dim\,$G$ be the generators of the
Lie group $G$ with structure constants $f^A_{BC}$ given by the commutation
relations
\begin{equation}\label{2.1}
[I_A, I_B]\=f^C_{AB}\, I_C \ .
\end{equation}
We normalize the generators such that the Killing-Cartan metric
on the Lie algebra $\gfrak$ of $G$ coincides with the Kronecker symbol,
\begin{equation}\label{2.2}
g_{AB}\=f^C_{AD}\,f^D_{CB}\=\de_{AB} \ .
\end{equation}
More general left-invariant metrics can be obtained by rescaling the
generators.

The Lie algebra $\gfrak$ of $G$ can be decomposed as
${\gfrak}={\hfrak}\oplus{\mfrak}$, where $\mfrak$ is the orthogonal complement
of the Lie algebra $\hfrak$ of $H$ in $\gfrak$. Then, the generators of $G$ can
be divided into two sets, $\{I_A\}=\{I_a\}\cup\{I_i\}$, where $\{I_i\}$ are 
the generators of $H$ with $i,j,\ldots = 1,\ldots,$\,dim\,$H$, 
and $\{I_a\}$ span the subspace $\mfrak$ of $\gfrak$, which can be identified 
with the tangent space at any given point $x\in G/H$ of the coset. 
For reductive homogeneous spaces we have the following commutation relations:
\begin{equation}\label{2.3}
[I_i, I_j]=f^k_{ij}\, I_k \ ,\quad [I_i, I_a]=f^b_{ia}\, I_b\ ,\quad
[I_a, I_b]=f^i_{ab}\, I_i + f^c_{ab}\, I_c\ .
\end{equation}
For the metric (\ref{2.2}) on $G$ we have
\begin{equation}\label{2.4}
g_{ij}\=f^k_{il}f^l_{kj} + f^b_{ia}f^a_{bj}\=\de_{ij}\ ,\quad g_{ia}\=0\ ,
\end{equation}
\begin{equation}\label{2.5}
g_{ab}\=2f^i_{ad}f^d_{ib} + f^c_{ad}f^d_{cb}\=\de_{ab}\ ,
\end{equation}
and more general metric can be obtained via rescaling of the generators.

\bigskip

\noindent
{\bf Spin connection on $G/H$.} 
On the group manifold $G$ we denote by $\hat E_A$ the left-invariant 
vector fields, which enjoy the same commutation relations as the matrices $I_A$
with $\{A\}=\{a,i\}$ in (\ref{2.3}). Let $\hat e^A$ be the left-invariant 
one-forms on $G$ dual to the vector fields $\hat E_A$. They satisfy 
the standard Maurer-Cartan equations. With the canonical projection
\begin{equation}\label{2.6}
\pi : \ G\to G/H
\end{equation}
we can push forward the $\hat E_A$ to vector fields $E_A=\pi_*\hat E_A$ 
on $G/H$ having the same commutation relations as $\hat E_A$.
Likewise, on $G/H$ we introduce left-invariant one-forms $e^A$ with
$\{A\}=\{a,i\}$ such that they are pulled back to $G$ via $\pi^*e^A=\hat e^A$.
These forms obey the Maurer-Cartan equations
\begin{equation}\label{2.7}
\diff e^a \= 
- f^a_{ib}\, e^i\wedge e^b -\sfrac{1}{2}\, f^a_{bc}\, e^b\wedge e^c
\und
\diff e^i \= 
- \sfrac12\,f^i_{bc}\, e^b\wedge e^c -\sfrac12\, f^i_{jk}\, e^j\wedge e^k \ ,
\end{equation}
which can be induced from $G$ or derived from the commutation relations for
the vector fields $E_a$ and $E_i$. On the other hand, the torsion-full
spin connection one-form $\om=(\om^a_b)$ is defined on $G/H$ by the equations
\begin{equation}\label{2.8}
\diff e^a + \om^a_{b}\wedge e^b\=T^a\ ,
\end{equation}
where
\begin{equation}\label{2.9}
T^a\=\sfrac{1}{2}\,T^a_{bc}\,e^b\wedge e^c
\end{equation}
and $T^a_{bc}$ are the components of the vector-valued torsion 
two-form $T=T^aE_a$.

We choose the torsion tensor components proportional to the structure constant
$f^a_{bc}$,
\begin{equation}\label{2.10}
T^a_{bc}\=\vk\,f^a_{bc}\ ,
\end{equation}
where $\vk$ is an arbitrary real parameter. 
Then, the affine spin connection on $G/H$ becomes
\begin{equation}\label{2.11}
\om^a_{b}\=f^a_{ib}e^i+\sfrac12\,(\vk{+}1)\,f^a_{cb}\,e^c\ =:\ \om^a_{cb}e^c
\end{equation}
with the coefficients
\begin{equation}\label{2.12}
\om^a_{cb}\=e^i_c\,f^a_{ib}+\sfrac{1}{2}\,(\vk{+}1)\,f^a_{cb}\ .
\end{equation}
In deriving (\ref{2.12}) from (\ref{2.7}) we used the fact that 
$g_{ab}=\de_{ab}$. For more details and applications see 
e.g.~\cite{KZ,Lust,MH}.

\bigskip

\noindent
{\bf Yang-Mills equations on $\R\times G/H$.}  
Consider the space $\R\times G/H$ with a coordinate $\tau$ on $\R$, 
a one-form $e^0:=\diff\tau$ and the metric
\begin{equation}\label{2.13}
\diff s^2 \= (e^0)^2 + \de_{ab}\,e^a e^b\ .
\end{equation}
For this direct product metric we have
\begin{equation}\label{2.14}
\om^0_{0b}\=\om^a_{0b}\=\om^0_{cb}\=0\ .
\end{equation}
Consider the principal bundle $P(\R{\times}G/H, G)$ over $\R\times G/H$ with
the structure group $G$ and a $\gfrak$-valued connection one-form $\Acal$ on
$P$ and the gauge field $\Fcal =\diff\Acal + \Acal\wedge\Acal$. In the basis
of one-forms $\{e^0, e^a\}$ on $\R\times G/H$, we have
\begin{equation}\label{2.15}
\Acal \=\Acal_0e^0 + \Acal_a e^a \und
\Fcal \=\Fcal_{0a}\,e^0\wedge e^a + \sfrac12\, \Fcal_{ab}\,e^a \wedge e^b\ .
\end{equation}
In the following we choose a `temporal' gauge in which 
$\Acal_0\equiv\Acal_\tau =0$.

The standard Yang-Mills equations on $\R\times G/H$ in components read
\begin{eqnarray}\label{2.16}
\nabla_a\Fcal^{a0}\ :=\ 
E_a\Fcal^{a0}+ \om^a_{ab}\Fcal^{b0}+[\Acal_a,\Fcal^{a0}] &=& 0\ , \\[6pt]
\label{2.17}
\nabla_0\Fcal^{0b}+\nabla_a\Fcal^{ab} \= 
E_0\Fcal^{0b}+E_a\Fcal^{ab}+ \om^d_{da}\Fcal^{ab}+ \om^b_{cd}\Fcal^{cd}
+[\Acal_a,\Fcal^{ab}] &=& 0\ ,
\end{eqnarray}
where we used (\ref{2.14}) and the gauge $\Acal_0=0$.

\bigskip

\noindent
{\bf Reduction to the kink equation.} 
We now make the $G$-equivariant ansatz (cf.~\cite{group6, popov1})
\begin{equation}\label{2.18}
\Acal \= e^iI_i+\p\,e^aI_a\qquad\Leftrightarrow\qquad
\Acal_a\=e^i_aI_i+\p\,I_a\ ,
\end{equation}
where $\p =\p (\tau )$ is a real function of $\tau\in\R$ only. 
The corresponding gauge field reads
\begin{equation}\label{2.19}
\begin{aligned}
\Fcal &\=\diff\Acal +\Acal\wedge\Acal\=\dot\p\,e^0\wedge e^aI_a-\sfrac{1}{2}\,
\left\{(1-\p^2)f^i_{bc}I_i + (\p -\p^2)f^a_{bc}I_a\right\}\,e^b\wedge
e^c\quad\Leftrightarrow\\[6pt]
\Fcal_{0a}&\=\dot\p\,I_a \und \Fcal_{bc}\=-\left\{(1-\p^2)f^i_{bc}I_i +
(\p -\p^2)f^a_{bc}I_a\right\}\ ,
\end{aligned}
\end{equation}
where a dot denotes a derivative with respect to $\tau$.

For our choice of the metric $g_{ab}=\de_{ab}$ and $g_{ij}=\de_{ij}$
one can pull down all indices in the Yang-Mills equations (\ref{2.16}) and
(\ref{2.17}). Substituting (\ref{2.18}) and (\ref{2.19}) reduces them to
\begin{equation}\label{2.20}
\begin{aligned}
\ddot\p\,I_a \ +\ & (1-\p^2)\,e^i_{c}I_j\,\left\{
f_{ibc}f_{jab}-f_{iab}f_{jbc}-f_{kac}f_{ijk} \right\}+\\
+\ & (\p-\p^2)\,e^i_{c}I_d\,\left\{
f_{ibc}f_{abd}-f_{iab}f_{bcd}+f_{acb}f_{ibd} \right\}+\\
+\ &\bigl[(\p-\sfrac{1}{2}(\vk{+}1))(\p-\p^2)f_{abc}f_{dbc} +
\p\,(1-\p^2)f_{aci}f_{dci}\bigr]I_d\=0\ .
\end{aligned}
\end{equation}
The expressions in the two braces vanish due to the Jacobi identity 
for the structure constants. To simplify (\ref{2.20}) further, 
we assume that the structure constants obey
\begin{equation}\label{2.21}
f_{aci}f_{bci}=\sfrac12(1{-}\a)\de_{ab}
\qquad\Leftrightarrow\qquad
f_{acd}f_{bcd}=\a\,\de_{ab}\ ,
\end{equation}
where we have taken into account (\ref{2.5}). 
This is true for a lot of homogeneous spaces, like e.g.~for
SU(3)/U(1)$^2$ and Sp(2)/Sp(1)$\times$U(1)~\cite{Lust, MH}.
In this case, (\ref{2.20}) reduces to the simple kink-type equation
\begin{equation}\label{2.22}
\begin{aligned}
2\,\ddot\p&\=(1{+}\a)\,\p^3-\a(\vk{+}3)\,\p^2-(1{-}\a(\vk{+}2))\,\p \\
&\=-\p\,(1-\p^2)\ +\ \a\,\p\,(1-\p)(2-\p)\ +\ \a\,\vk\,\p\,(1-\p) \\
&\=(1{+}\a)\,\p\,(\p-1)\,\bigl(\p-\sfrac{(\vk+2)\a-1}{\a+1}\bigr) \ 
=:\ V'(\p)\ .
\end{aligned}
\end{equation}
which is the static equation of motion for a $\p^4$-model with potential~$V$
in 1+1 dimensions.

\bigskip

\section{Yang-Mills solutions on symmetric spaces}

\noindent
{\bf BPS kink equations for $\a =0$.} 
Let $G/H$ be a symmetric space,
i.e.~$f_{abc}{=}0$ and the geometric torsion~(\ref{2.10}) vanishes, so that
(\ref{2.11}) becomes the Levi-Civita connection on $G/H$. Formally, this case
can be obtained by choosing $\a{=}0$ in~(\ref{2.21}), which eliminates $\vk$ 
and simplifies (\ref{2.22}) to\footnote{
Another special choice is $\a{=}1$, corresponding to $G/H$ being a group,
i.e.~$H=\{1\}$. Taking $\vk{=}0$ then yields 
$2\ddot\p=\p(1{-}\p)(1{-}2\p)=-2[\phi{-}\sfrac12](\sfrac14-[\p{-}\sfrac12]^2)$,
which integrates to $2\dot\p=\sqrt{2}(\sfrac14-[\p{-}\sfrac12]^2)$
and also describes a kink.
}
\begin{equation}\label{3.1}
\ddot\phi \= -\sfrac{1}{2}\,\phi\,(1-\phi^2)\ .
\end{equation}
Furthermore, when we substitute the
ansatz (\ref{2.18}) into the Yang-Mills action functional
\begin{equation}\label{3.2}
S\=-\sfrac{1}{4}\,\int_{\R\times G/H}\tr\, (\Fcal\wedge *\Fcal )\ ,
\end{equation}
it reduces to a factor proportional to Vol($G/H$) times the
$\p^4$ energy functional 
\begin{equation}\label{3.3}
\begin{aligned}
E&\=\int\limits_{-\infty}^{\infty}\diff \tau\,\left\{\dot\phi^2
+ \sfrac{1}{4}(1 -\phi^2)^2\right\} \= \\
&\=\int\limits_{-\infty}^{\infty}\diff \tau\,\left (\dot\phi\mp
\sfrac{1}{2}(1 -\phi^2)\right )^2\pm \int\limits_{-\infty}^{\infty}
\diff \phi\;(1 -\phi^2)\ \ge\ \sfrac{4}{3}\,|q|\ ,
\end{aligned}
\end{equation}
where
\begin{equation}\label{3.4}
q\ :=\ \sfrac{1}{2}\int\limits_{-\infty}^{\infty}\diff \tau\,\dot\phi \=
\sfrac{1}{2}\,\bigl(\phi (+\infty) -\phi (-\infty)\bigr)\ \in\{1,0,-1\}
\end{equation}
is the topological charge (see e.g.~\cite{MS}). 
Here, $\p(+\infty){=}\pm 1$ and $\p(-\infty){=}\mp 1$
are the vacua of the kink model~(\ref{3.3}). 

The inequality in (\ref{3.3}) is saturated on the BPS kink equations
\begin{equation}\label{3.5}
\dot\phi \= \pm\sfrac{1}{2} (1-\phi^2)\ ,
\end{equation}
where the sign choice corresponds to the sign of $q$.
For the $+$ sign, the solution is known as the $\p^4$ kink,
\begin{equation}\label{3.6}
\phi \=\tanh\sfrac{\tau}{2}\ ,
\end{equation}
which interpolates between the vacua at $-1$ and $+1$ and 
carries the topological charge $q{=}1$. For $q{=}-1$, the antikink 
\begin{equation}\label{3.7}
\phi \= -\tanh\sfrac{\tau}{2}
\end{equation}
describes a transition from $+1$ to $-1$. Both the static kink and antikink
possess an energy of $E{=}\frac43$ (cf.~(\ref{3.3})).

\noindent
{\bf Instantons on $\R\times G/H$.}
We now turn to the asymptotic behaviour of the gauge potential
\begin{equation}\label{3.8}
\Acal \= e^iI_i\ +\ \tanh\sfrac{\tau}{2}\, e^aI_a\ ,
\end{equation}
which by construction solves the Yang-Mills equations on $\R\times G/H$. 
There, `infinity' is the disconnected manifold
\begin{equation}\label{3.9}
G/H\times \Z_2 \=(G/H)_- \cup\ (G/H)_+\ ,
\end{equation}
where $(G/H)_{\pm}=G/H\times\{\tau{=}\pm\infty\}$. Introducing the notation
$\Acal_{\pm}:=\Acal(\tau{=}\pm\infty)$, from (\ref{3.8}) and
\begin{equation}\label{3.10}
\Fcal \= \sfrac12 \cosh^{-2}\!\sfrac{\tau}{2}\;
\bigl( e^0\wedge e^a\, I_a - f^i_{ab}\, e^a\wedge e^b\, I_i \bigr)
\end{equation}
we obtain
\begin{equation}\label{3.11}
\Acal_-= h^{-1}_-\diff h_-\und \Acal_+= h^{-1}_+\diff h_+\qquad\Leftrightarrow
\qquad\Fcal_\pm =0 \quad\textrm{(vacua)}\ ,
\end{equation}
where $h_{\pm}$ are $G$-valued functions. One can employ $h_-$ to gauge
transform $\Acal_-$ to zero, i.e.~pass to
\begin{equation}\label{3.12}
\tilde\Acal \= h_-\Acal h^{-1}_- + h_-\diff h^{-1}_-
\qquad\Leftrightarrow\qquad
\tilde\Acal_-\=0\und\tilde\Acal_+\=g^{-1}\diff g\ ,
\end{equation}
where the map
\begin{equation}\label{3.13}
g\=h_+h^{-1}_- : G/H \to G
\end{equation}
has degree one equal to the topological charge $q{=}1$ of the kink~$\p$. 
Thus, the solution (\ref{3.8}) to the Yang-Mills equations on
$\R\times G/H$ is an instanton configuration describing a transition between
the vacua $\tilde\Acal_-{=}0$ and $\tilde\Acal_+{=}g^{-1}\diff g$. The action
functional on this solution is finite as was shown above.

\noindent
{\bf Instanton-anti-instanton chains.} 
One may obtain a different kind of solutions (sphalerons) to~(\ref{3.1}) 
by imposing the periodic boundary conditions
\begin{equation}\label{3.14}
\phi (\tau{+}L)\=\phi (\tau)\ .
\end{equation}
Such a configuration can be considered as a function on a circle $S^1$ 
with circumference $L$. Sphalerons are given by~\cite{group7}
\begin{equation}\label{3.15}
\phi(\tau ; k)\=2k\,b(k)\;\sn[b(k)\tau;k]\qquad \textrm{with}\qquad
b(k)\=(2+2k^2)^{-1/2}\quad \textrm{and}\quad 0\le k\le 1\ .
\end{equation}
Since the Jacobi elliptic function $\sn[u;k]$ has a period of $4{\cal K}(k)$,
the condition (\ref{3.14}) is satisfied if
\begin{equation}\label{3.16}
b(k)\,L\=4{\cal K}(k)\,n \qquad\textrm{for}\quad n\in\NN\ ,
\end{equation}
which fixes $k=k(L,n)$ so that $\phi(\tau;k)=:\phi_n(\tau)$.
Solutions (\ref{3.15}) exist if $L\ge L_n:=2\pi\sqrt{2}\,n$~\cite{group7}.

By virtue of the periodic boundary condition (\ref{3.14}), 
the topological charge of the sphaleron (\ref{3.15}) is zero. 
In fact, the configuration (\ref{3.15}) is interpreted as a chain 
of $n$ kinks and $n$ antikinks, alternating and equally spaced 
around the circle~\cite{MS, group7}. With $n{=}1$ for instance,
we encounter a single kink-antikink pair winding once around~$S^1$.
The energy (\ref{3.3}) of the sphaleron (\ref{3.15}) is
\begin{equation}\label{3.17}
E[\phi_n]\=\frac{2n}{3\sqrt{2}}\left [8(1+k^2)\,\Ecal(k)\ -\
(1-k^2)(5+3k^2)\,{\cal K}(k)\right ]\ ,
\end{equation}
where ${\cal K}(k)$ and $\Ecal(k)$ are the complete elliptic integrals of 
the first and second kind, respectively~\cite{group7}.
In the limit $L\to\infty$ one approaches a superposition of kinks~(\ref{3.6})
and antikinks~(\ref{3.7}), and the energy becomes additive,
$E[\phi_n]\to2n\cdot\frac43$.

Substituting the non-BPS solution (\ref{3.15}) of (\ref{3.1}) 
into (\ref{2.18}) and (\ref{2.19}), we obtain a finite-action configuration
\begin{equation}\label{3.18}
\Acal \= e^iI_i + \p_n\, e^aI_a\ ,\qquad
\Fcal \= \dot\p_n\,e^0\wedge e^a\, I_a - \sfrac{1}{2}\left \{
(1-\p_n^2)f^i_{bc} I_i + (\p_n-\p_n^2)f^a_{bc} I_a\right \}\, e^b\wedge e^c\ ,
\end{equation}
which is interpreted as a chain of $n$ instanton-anti-instanton pairs 
sitting on $S^1_L\times G/H$.

\bigskip

\section{Yang-Mills solutions on nonsymmetric coset spaces}

\noindent
{\bf BPS kink equations for $\vk\not=0$.}  
On nonsymmetric coset spaces we need the torsional freedom in order
to find BPS solutions to the kink-type equation~(\ref{2.22}).
For a given value of $\a\ne0$ it can be reduced to the standard static form 
of the kink equation by suitably adjusting the parameter $\vk$ and 
by shifting the field via
\begin{equation}\label{4.1}
\p\=\vp+\b
\end{equation}
in such a way that the $O(\vp^0)$ and $O(\vp^2)$ terms vanish in~(\ref{2.22}).
This is achieved for
\begin{equation}\label{4.2}
\b \= \frac{\vk{+}3}{3}\,\frac{\a}{1{+}\a} \und
\vk=-3\ ,\quad\vk=\frac{3(1{-}\a)}{2\a}\quad\textrm{or}\quad\vk=\frac3{\a} \ .
\end{equation}
Furthermore, it is convenient to rescale $\tau\to\tau/\sqrt{1{+}\a}$.
For the three special values of $(\b,\vk)$ we then obtain the equations
and corresponding scalar potentials
\alpheqn
\begin{align}
\textrm{a)}\quad (\b,\vk)&=(0,-3):\quad& 
2\ddot\p&\=-\p\,(1-\p^2) 
\quad&\Leftrightarrow\quad 
V&\=\sfrac14\bigl(\p^2-1\bigr)^2 
\label{4.3a}\\[6pt]
\textrm{b)}\quad (\b,\vk)&=(\sfrac12,\sfrac{3(1{-}\a)}{2\a}):\quad&
2\ddot\p&\=-[\p{-}\sfrac12](\sfrac14-[\p{-}\sfrac12]^2)
\quad&\Leftrightarrow\quad 
V&\=\sfrac14\bigl([\p{-}\sfrac12]^2-\sfrac14\bigr)^2 
\label{4.3b}\\[6pt]
\textrm{c)}\quad (\b,\vk)&=(1,\sfrac3{\a}):\quad&
2\ddot\p&\=-[\p{-}1](1-[\p{-}1]^2)
\quad&\Leftrightarrow\quad 
V&\=\sfrac14\bigl([\p{-}1]^2-1\bigr)^2\ .
\label{4.3c}
\end{align}
\reseteqn

By construction, these equations of motion can be integrated to 
first-order (BPS) equations $\dot\p=\pm\sqrt{V}$ with well known solutions,
\alpheqn
\begin{align}
\textrm{a)}\qquad 2\dot\p&\=\pm(1-\p^2) 
&\Leftrightarrow\qquad
\p&\=\pm\tanh\sfrac{\tau}2 \label{4.4a}\\[6pt]
\textrm{b)}\qquad 2\dot\p&\=\pm(\sfrac14-[\p{-}\sfrac12]^2) 
&\Leftrightarrow\qquad
\p&\=\sfrac12\pm\sfrac12\tanh\sfrac{\tau}4 \label{4.4b}\\[6pt]
\textrm{c)}\qquad 2\dot\p&\=\pm(1-[\p{-}1]^2) 
&\Leftrightarrow\qquad
\p&\=1\pm\tanh\sfrac{\tau}2\ , \qquad\qquad\qquad{}\label{4.4c}
\end{align}
\reseteqn
where we chose the integration constant~$\tau_0$ in a symmetric manner.
The three solutions may be described simultaneously by as
\begin{equation}\label{4.5}
\p\=\b\pm\g\tanh\sfrac{\g\tau}{2} \qquad\textrm{with}\quad \g=1,\sfrac12,1\ ,
\end{equation}
respectively. For the upper sign choice, these solutions are $\p^4$ kinks 
with a topological charge of $q=+1$, which interpolate between 
the vacua $\p(-\infty)=\b{-}\g$ and $\p(+\infty)=\b{+}\g$ 
via the critical point $\p(0)=\b$.
The lower sign choice corresponds to the antikinks and $q=-1$.
The critical points $\ph$ of the $\p^4$ potentials~$V$ in the three cases are
\begin{equation}\label{4.6}
\begin{array}{cccccc}
\qquad\textrm{case}\qquad & \quad\b\quad & \quad\g\quad &
\quad\p(-\infty)\quad & \quad\p(0)\quad & \quad\p(+\infty)\quad \\[4pt]
\textrm{a)}   &     0    &     1    &  -1         &     0    &   1         \\
\textrm{b)}   & \sfrac12 & \sfrac12 &   0         & \sfrac12 &   1         \\
\textrm{c)}   &     1    &     1    &   0         &     1    &   2         
\end{array} \ .
\end{equation}


\noindent
{\bf Modifications of bundles over $G/H$.} 
Inserting (\ref{4.5}) into (\ref{2.18}) and (\ref{2.19}), we obtain 
the corresponding solutions to the Yang-Mills equations on $\R\times G/H$. 
For the upper sign choice they read
\begin{equation}\label{4.7}
\begin{aligned}
\Acal &\= e^iI_i + (\b+\g\tanh\sfrac{\g\tau}{2})\,e^aI_a \und \\[6pt]
\Fcal &\= \sfrac{\g^2}{2}\cosh^{-2}\!\sfrac{\g\tau}{2}\,e^0{\wedge}\,e^a\,I_a\
-\ \sfrac12\bigl\{h(\tau)\,f^i_{bc}I_i + m(\tau)\,f^a_{bc}I_a\bigr\}\,
e^b{\wedge}\,e^c\ ,
\end{aligned}
\end{equation}
where we introduced the functions
\begin{equation}\label{4.8}
\begin{aligned}
h(\tau)&\=(1-\b-\g\tanh\sfrac{\g\tau}{2})\,(1+\b+\g\tanh\sfrac{\g\tau}{2})\ ,\\
m(\tau)&\=(1-\b-\g\tanh\sfrac{\g\tau}{2})\,(\b+\g\tanh\sfrac{\g\tau}{2})\ .
\end{aligned}
\end{equation}
Due to the asymptotic behavior of $h(\tau)$ and $m(\tau)$,
the standard Yang-Mills action functional on these solutions is infinite,
in contrast to the symmetric case~(\ref{3.10}).
However, for any fixed $\tau$, including $\tau=\pm\infty$, the action
reduced to the compact coset $G/H$ is finite.

In each case a)--c), the Yang-Mills configuration (\ref{4.7}) describes 
a transition between the three critical points $\ph$ with $V'(\ph)=0$ 
at $\tau{=}-\infty$, $\tau{=}0$ and $\tau{=}+\infty$. 
These critical points are characterized by $\ddot\p=0=\ddot\Acal$, 
and hence the gauge fields there satisfy the Yang-Mills equations on $G/H$:
\begin{equation}\label{4.9}
\Acal_{\rm crit} \= e^iI_i + \ph\,e^aI_a \und
\Fcal_{\rm crit} \= 
-\sfrac12\bigl\{(1-\ph)(1+\ph)\,f^i_{bc}I_i
               +(1-\ph)\,\ph\,f^a_{bc}I_a \bigr\}\,e^b{\wedge}\,e^c\ .
\end{equation}

Recall that we consider the principal bundle $P(\R\times G/H, G)$ over
$\R\times G/H$ with the structure group $G$. Given a finite-dimensional
representation $V_G$ of the group $G$, the corresponding associated vector
bundle over $\R\times G/H$ is given by the fibred product
\begin{equation}\label{4.10}
\Ecal \= P(\R\times G/H, G)\times_G\, V_G\ \to\ \R\times G/H\ .
\end{equation}
Our connection $\Acal$ is defined on $\Ecal$. 
However, for any given $\tau\in\R$, one can restrict the bundle 
$\Ecal\to\R\times G/H$ to the bundle $\Ecal_{\tau}\to G/H$. 
In particular, the connections $\Acal_{\rm crit}$ are defined on bundles 
$\Ecal_{\rm crit}$ over $G/H$ and satisfy the Yang-Mills equations on~$G/H$.
In our three cases a)--c), we encounter five such bundles:
\begin{equation}\label{4.11}
\begin{array}{lccccc}
\textrm{value of $\ph$} & -1 & 0 & 1/2 & 1 & 2 \\
\textrm{bundle}\ \Ecal_{\rm crit} & \Ecal_- & \Ecal_{\rm can} & \Ecal_0 & 
\Ecal_{\rm flat} & \Ecal_+ \\
\textrm{topology}\qquad&\textrm{irreducible}&\textrm{reducible}&
\textrm{irreducible}&\textrm{trivial}&\textrm{irreducible}\\
\hat{h}\;=1{-}\ph^2 &  0 & 1 & 3/4 & 0 & -3 \\
\hat{m}=\ph{-}\ph^2 & -2 & 0 & 1/4 & 0 & -2 
\end{array}\ .
\end{equation}
The bundles $\Ecal_-$ and $\Ecal_+$ are irreducible and topologically 
nontrivial, $\Ecal_{\rm can}$ is topologically nontrivial but 
reducible,\footnote{
Its fibres split into a direct sum of irreducible representations 
of the group $H$.}
and the bundle $\Ecal_{\rm flat}=G/H\times V_G$ is trivial.

The configurations (\ref{4.7}) then correspond to the following 
modifications of non-Abelian bundles over $G/H$ (cf.~\cite{KW, LOZ}),
\begin{equation}\label{4.12}
\Ecal_-\ \to\ \Ecal_{\rm can}\ \to\ \Ecal_{\rm flat} \qquad\textrm{or}\qquad
\Ecal_{\rm can}\ \to\ \Ecal_0\ \to\ \Ecal_{\rm flat} \qquad\textrm{or}\qquad
\Ecal_{\rm can}\ \to\ \Ecal_{\rm flat}\ \to\ \Ecal_+ \ ,
\end{equation}
and we see that the connections given in (\ref{4.7}) describe interpolations 
between bundles of all types. 
The curvatures of the critical connections are given by~(\ref{4.9}) as
\begin{equation}\label{4.13}
\Fcal_{\rm crit} \= -\sfrac12\bigl\{
\hat{h}\,f^i_{bc}I_i + \hat{m}\,f^a_{bc}I_a \bigr\}\,e^b{\wedge}\,e^c\ ,
\end{equation}
with the values of $\hat{h}$ and $\hat{m}$ displayed in~(\ref{4.11}).
We note that the case of symmetric coset spaces, treated in the previous
section, may be included here by adding
\begin{equation}\label{4.14}
\textrm{case d)} \qquad \b=0\ ,\quad \g=1\ ,\quad m(\tau)\equiv0\ ,\quad
h(\tau)=\cosh^{-2}\!\sfrac{\tau}{2}\ ,\quad 
\Ecal_{\rm flat}\ \to\ \Ecal_{\rm can}\ \to\ \Ecal_{\rm flat}\ . 
\end{equation}
It resembles case a), but in (\ref{4.11}) the bundle $\Ecal_-$ at 
$\hat{\p}{=}-1$ gets replaced by $\Ecal_{\rm flat}$ due to $\hat{m}{=}0$.

\bigskip

\noindent
{\bf Chains of bundle modifications.} 
One can also reduce the Yang-Mills equations on the nonsymmetric coset spaces 
$S^1\times G/H$ and obtain sphaleron solutions analogous to (\ref{3.15}) 
with (\ref{3.16}). Then substituting them into (\ref{2.18}) and (\ref{2.19}), 
we obtain configurations which describe chains of bundle modifications 
followed by the inverse bundle modifications. The action functional on
these configurations is finite for $G/H$ being compact.

\bigskip

\noindent
{\bf Dyons on $\R\times G/H$.} 
Let us finally change the signature of the metric on
$\R\times G/H$ from Euclidean to Lorentzian by choosing on $\R$ a
coordinate $t=-\im\tau$  so that $\tilde e^0 = \diff t= -\im\diff\tau$.
Then as metric on $\R\times G/H$ we have
\begin{equation}\label{4.15}
\diff s^2 \= -(\tilde e^0)^2 +\de_{ab}e^ae^b\ .
\end{equation}
For $\Acal$ we copy the ansatz (\ref{2.18}) and obtain
\begin{equation}\label{4.16}
\Fcal\=\sfrac{\diff\p}{\diff t}\,\tilde e^0{\wedge}\,e^a I_a -
\sfrac12\bigl\{(1-\p^2)f^i_{bc}I_i + (\p-\p^2)f^a_{bc}I_a\bigr\}
\,e^b{\wedge}\,e^c\ .
\end{equation}
After substituting (\ref{2.18}) and (\ref{4.16}) into the Yang-Mills
equations on $\R\times G/H$, we arrive at the same second-order differental
equations as in the Euclidean case, except for the replacement
\begin{equation}\label{4.17}
\ddot\p \quad\longrightarrow\quad -\sfrac{\diff^2\p}{\diff t^2}
\qquad\textrm{and hence}\qquad V\quad\longrightarrow\quad -V\ .
\end{equation}
In particular, this implies a sign change of the left-hand side
relative to the right-hand side in (\ref{2.22}), (\ref{3.1}) and~(4.3).
Although the Lorentzian equations do not follow from first-order equations,
they can still be integrated explicitly,
\begin{equation}\label{4.18}
\phi(t) \= \b + \sqrt{2}\,\g\,\cosh^{-1}\!\sfrac{\g t}{\sqrt{2}}\ ,
\end{equation}
again after rescaling $t\to t/\sqrt{1{+}\a}$ and with the $(\b,\g)$ values
from~(\ref{4.6}).
This configuration is the bounce in an inverted double-well potential,
which from the asymptotical local minimum briefly explores the unstable region.

Inserting (\ref{4.18}) into (\ref{2.18}) and (\ref{2.19}), we arrive at
dyon-type configurations with smooth nonvanishing `electric' and `magnetic'
field strengths $\Fcal_{0a}$ and $\Fcal_{ab}$, respectively.
For instance, on a symmetric coset we have $(\b,\g){=}(0,1)$ and get
\begin{equation}\label{4.19}
\begin{aligned}
\Acal &\= e^iI_i + \sqrt{2}\cosh^{-1}\!\sfrac{t}{\sqrt{2}}\,e^aI_a \und\\[6pt]
\Fcal &\=-\sfrac12\cosh^{-2}\!\sfrac{t}{\sqrt{2}}\,
\bigl( 2\sinh\sfrac{t}{\sqrt{2}}\ \diff t\,{\wedge}\,e^a\,I_a +
(\sinh^2\!\sfrac{t}{\sqrt{2}}-1)\, f^i_{bc} I_i\, e^b{\wedge}\,e^c\bigr)\ .
\end{aligned}
\end{equation}
The full energy 
$-\tr(2\Fcal_{0a}\Fcal_{0a}+\Fcal_{ab}\Fcal_{ab})\times{\rm Vol}(G/H)$ 
for this configuration (\ref{4.19}) is finite, but the action is not.
The same is true for the nonsymmetric space configurations, obtained by 
substituting (\ref{4.18}) into (\ref{2.18}) and (\ref{2.19}).

\section{Yang-Mills flows on $G/H$}

\noindent
{\bf Reduction of the Yang-Mills flow equations.} 
Consider the Yang-Mills flow on
a reductive coset space $G/H$ which is a one-parameter family of
connections\footnote{These $\Acal (\tau )$ with $\tau\in\R$ can be
viewed either as connections on a family $\Ecal_\tau$ of vector bundles
over $G/H$ or as a connection $\Acal =\{\Acal (\tau )\}$ on a vector
bundle $\Ecal$ over $\R\times G/H$.} $\Acal (\tau)$, determined by the
differential equation (\ref{1.3}). In the left-invariant basis $\{e^a\}$
of one-forms on $G/H$ we can rewrite (\ref{1.3}) in components as
\begin{equation}\label{5.1}
\dot\Acal_a \= -\nabla_b\Fcal_{ab}\ .
\end{equation}
Recall that we have chosen the structure constants of $G$ such that the metric 
on $G/H$ has components $\delta_{ab}$ and hence we can pull all indices down.

Again we employ the $G$-equivariant ansatz (\ref{2.18}) for $\Acal_a$,
\begin{equation}\label{5.2}
\Acal_a \= e_a^i\,I_i+\p\,I_a\qquad\Rightarrow\qquad
\Fcal_{ab}\=-\left\{(1-\p^2)f^i_{ab}I_i + (\p -\p^2)f^c_{ab}I_c\right\} \ .
\end{equation}
Substituting (\ref{5.2}) into (\ref{5.1}) yields the equation
\begin{equation}\label{5.3}
\begin{aligned}
2\,\dot\p&\=(1{+}\a)\,\p^3-\a(\vk{+}3)\,\p^2-(1{-}\a(\vk{+}2))\,\p \\
&\=-\p\,(1-\p^2)\ +\ \a\,\p\,(1-\p)(2-\p)\ +\ \a\,\vk\,\p\,(1-\p) \\
&\=(1{+}\a)\,\p\,(\p-1)\,\bigl(\p-\sfrac{(\vk+2)\a-1}{\a+1}\bigr)\=V'(\p)\ ,
\end{aligned}
\end{equation}
which differs from (\ref{2.22}) only by the order of the derivative of the 
left-hand side.
The stable points of this flow, $\dot\p{=}0$, are again precisely 
the critical points of the $\p^4$ potential, 
\begin{equation}\label{5.4}
V'(\hat{\p})=0 \qquad\Leftrightarrow\qquad
\hat{\p}=0\ ,\quad \hat{\p}=1\quad\textrm{and}\quad
\hat{\p}=\sfrac{(\vk+2)\a-1}{\a+1}\ =:\rho\ .
\end{equation}
While the first two $\hat{\p}$ values produce just the canonical bundle 
$\Ecal_{\rm can}$ and the flat bundle $\Ecal_{\rm flat}$, respectively, 
with the connection and field strength given in~(\ref{4.9}), $\hat{\p}{=}\rho$
depends on $\vk$ and $\a$ and may take any real value. This produces a family 
of solutions to the Yang-Mills equations on $G/H$, with field strengths
\begin{equation}\label{5.5}
\Fcal_{\rm crit}(\a,\vk) \= 
-\sfrac{1}{2(\a+1)^2}\Bigl\{\bigl(2-[\vk{+}1]\a\bigr)[\vk{+}3]\a\,f^i_{bc}I_i
+\bigl(2-[\vk{+}1]\a\bigr)\bigl([\vk{+}2]\a-1\bigr)\,f^a_{bc}I_a \Bigr\}\,
e^b{\wedge}\,e^c\ .
\end{equation}
In general, the corresponding $\p^4$ potential is asymmetric and not of
BPS type. Only for the special values given in~(\ref{4.2}), the potential
minima are degenerate, and the BPS configurations discussed in section~4
are recovered. The three cases a)--c) of section~4 simply correspond to the 
situations where $\rho$ lies to the left, in between, or to the right of 
the two fixed critical points, respectively.
For symmetric coset spaces, i.e.~case d), $\a{=}0{=}f_{abc}$ reproduces
$\Fcal_{\rm crit}=\Fcal_{\rm flat}=0$.

\bigskip

\noindent
{\bf Explicit solutions.} 
The integration of~(\ref{5.3}) yields
\begin{equation}\label{5.6}
\p^{\rho-1}\,(\p{-}1)^{-\rho}\;(\p{-}\rho) \= 
C\,\exp\bigl\{\sfrac12\rho(\rho{-}1)\,(\a{+}1)\,\tau\bigr\}
\end{equation}
with an integration constant~$C$, as long as $\rho\neq0$ or $1$.
Let us absorb $\a$ by rescaling $\tau\to\tau/(1{+}\a)$.
The implicit solution~(\ref{5.6}) simplifies in the four special cases a)--d):
\begin{equation}\label{5.7}
\begin{array}{cllcl}
\textrm{a)} \quad & \vk=-3\ ,\quad                   & \rho=-1       & 
\qquad\Rightarrow\qquad & 
\p\=\pm(1+\exp\tau)^{-1/2} \\[6pt]
\textrm{b)} \quad & \vk=\sfrac{3(1-\a)}{2\a}\ ,\quad & \rho=\sfrac12 & 
\qquad\Rightarrow\qquad & 
\p\=\sfrac12\pm\sfrac12(1+\exp\sfrac{\tau}{4})^{-1/2}\\[6pt]
\textrm{c)} \quad & \vk=\sfrac{3}{\a}\ ,\quad        & \rho=2        & 
\qquad\Rightarrow\qquad & 
\p\=1\pm(1+\exp\tau)^{-1/2} \\[6pt]
\textrm{d)} \quad & \a=0\ ,\quad                     & \rho=-1       & 
\qquad\Rightarrow\qquad & 
\p\=\pm(1+\exp\tau)^{-1/2}
\end{array} \ ,
\end{equation}
where we have chosen $C{=}{-}1$ or $C^2{=}{-}1$ for simplicity.
Choosing the lower sign above, all these configurations interpolate between
the left minimum of the potential, $\p(-\infty)=\b{-}\g$, and the unstable
middle extremum, $\p(+\infty)=\b$.
This implies that the flow of the Yang-Mills potentials and field strengths
covers just half of the transition path of the bundle modifications in
(\ref{4.12}) and~(\ref{4.14}).
Apart from this difference, we see that the Yang-Mills flow equations
(\ref{5.1}) produce the same type of bundle modifications as the solutions
to the second-order Yang-Mills equations on~$\R\times G/H$.

\bigskip

\noindent
{\bf First-order flow equations.} 
Recall that our ansatz (\ref{2.18}) reduces
the Yang-Mills equations on $\R\times G/H$ to the ordinary second-order 
differential equation (\ref{2.22}) for~$\p$. Its kink and antikink solutions,
however, obey the first-order (BPS) equation (\ref{3.5}) or~(4.4).
Therefore, it is reasonable to search for a BPS analogue of the Yang-Mills 
flow equations~(\ref{5.1}), which would be algebraic in~$\Fcal$ on the
right-hand side. A well known example are the Yang-Mills self-duality equations
in temporal gauge on $\R\times G/H = \R\times S^3$,
\begin{equation}\label{5.8}
\dot\Acal _a \= - \sfrac{1}{2}\,\eps_{abc}\,\Fcal_{bc}\ ,
\end{equation}
where $a,b,\ldots=1,2,3$ and $\eps_{abc}$ are the structure constants of SU(2).
This case was generalized in~\cite{IvLe} to manifolds $\R\times G$, 
where $G$ is a semisimple Lie group, and gave an analogue of the generalized 
self-duality equations on~$\R^d$ introduced in~\cite{CDFN, W1}. 
In all these cases the full Yang-Mills equations follow
from the first-order BPS equations.

Here we introduce the first-order flow equations
\begin{equation}\label{5.9}
\dot\Acal _a \= \mp {\la}\,f_{abc}\,\Fcal_{bc}\ ,
\end{equation}
where $f_{abc}$ form the part of the structure constants of~$G$ related 
to the subspace~$\mfrak\subset\gfrak$, as given in~(\ref{2.3}).
Hence, this flow is trivial for symmetric coset spaces.
Note that stable points $\dot\Acal=0$ of these flow equations are given by
\begin{equation}\label{5.10}
f_{abc}\,\Fcal_{bc}\=0\ ,
\end{equation}
which is in fact true both for the canonical and for the flat connection
on (a bundle over)~$G/H$.
Solutions to (\ref{5.9}) do not necessarily satisfy the standard Yang-Mills 
flow equations on $G/H$ or the full Yang-Mills equations\footnote{
However, this can happen in some special cases, e.g.~for Spin(7)$/G_2 = S^7$.}
on~$\R\times G/H$. In this sense (\ref{5.9}) are similar to the $B_n$
and $C_n$ BPS-type gauge equations considered in~\cite{W1}. Yet, for our
equivariant ansatz (\ref{2.18}) the first-order flow equations~(\ref{5.9}) 
reduce to the equations
\begin{equation}\label{5.11}
\dot\p \= \pm\la\,(\sfrac14-[\p{-}\sfrac12]^2)\ ,
\end{equation}
which coincides with (\ref{4.4b}). Thus, in this case the kink equations
do descend from the BPS-type gauge equations~(\ref{5.9}), and the kink
of~(\ref{4.4b}) yields a solution~(\ref{4.7}) to~(\ref{5.9}) which also 
solves the full Yang-Mills equations on~$\R\times G/H$.

\bigskip

\noindent
{\bf Acknowledgements}

\medskip

\noindent
This work was supported in part by the cluster of excellence EXC 201 
``Quantum Engineering and Space-Time Research'',
by the Deutsche Forschungsgemeinschaft (DFG)
and by the Heisenberg-Landau program.
The work of T.A.I. and A.D.P. was partially supported by the 
Russian Foundation for Basic Research (grant RFBR 09-02-91347).

\end{document}